# Predicting Compressive Strength of Consolidated Molecular Solids Using Computer Vision and Deep Learning

Brian Gallagher[a], Matthew Rever[b], Donald Loveland[c], T. Nathan Mundhenk[b], Brock Beauchamp[b], Emily Robertson[c], Golam G. Jaman[d], Anna M. Hiszpanski[c], and T. Yong-Jin Han*[c]

[a]Center for Applied Scientific Computing, Lawrence Livermore National Laboratory
[b]Computational Engineering Division, Lawrence Livermore National Laboratory
[c]Materials Science Division, Lawrence Livermore National Laboratory
[d]Department of Electrical Engineering, Idaho State University

Corresponding Author: T. Yong-Jin Han, han5@llnl.gov

## Abstract
We explore the application of computer vision and machine learning (ML) techniques to predict material properties (e.g., compressive strength) based on SEM images. We show that it's possible to train ML models to predict materials performance based on SEM images alone, demonstrating this capability on the real-world problem of predicting uniaxially compressed peak stress of consolidated molecular solids samples. Our image-based ML approach reduces mean absolute percentage error (MAPE) by an average of 24% over baselines representative of the current state-of-the-practice (i.e., domain-expert's analysis and correlation). We compared two complementary approaches to this problem: (1) a traditional ML approach, random forest (RF), using state-of-the-art computer vision features and (2) an end-to-end deep learning (DL) approach, where features are learned automatically from raw images. We demonstrate the complementarity of these approaches, showing that RF performs best in the "small data" regime in which many real-world scientific applications reside (up to 24% lower RMSE than DL), whereas DL outpaces RF in the "big data" regime, where abundant training samples are available (up to 24% lower RMSE than RF). Finally, we demonstrate that models trained using machine learning techniques are capable of discovering and utilizing informative crystal attributes previously underutilized by domain experts.

## 1. Introduction

### 1.1 Background
Materials characterization is a cornerstone in materials science, providing insights on materials' structures and properties to further the understanding of fundamental phenomena and guide the materials optimization process for application development. Visualization techniques, such as scanning and transmission electron microscopy, electron diffraction, X-ray computed tomography and magnetic resonance imaging, among others, are widely used providing high spatial resolution images of atomic arrangements, morphologies, particle shapes, and microstructure information including defects and voids within materials. With significant improvements in science and technology of materials characterization methods, visualization tools listed above have also made advancements providing higher resolutions and faster data collection capabilities. With these breakthroughs in visualization techniques, the bottleneck in advancements in materials characterization will no longer be the capability limitations of the characterization tools themselves, but rather the ability to rapidly analyze and interpret the large amount of high-quality data.

Additionally, one of the most beneficial aspects of visualization characterization techniques is the immediate feedback one receives upon analyzing one's samples. For example, a scanning electron microscopy with Energy Dispersive Spectroscopy (SEM-EDS) provides its users immediate information regarding size, morphology, composition and other microstructure information. However, as samples become more complex and heterogeneous, the immediate feedback is no longer definitive, as information becomes *qualitative*, and not *quantitative*. To obtain more quantitative values, additional analyses are needed, especially for heterogeneous samples. As images or micrographs collected by SEM and other visualization tools become more complex (high dimensional data), additional analysis methods are needed to impart significance to the data.  Although human interpretation is often sufficient to elucidate the significance of the visual data, it also can introduce personal bias, which can overlook or neglect potentially important information.

To reduce human workload and to accelerate extraction of quantifiable values from SEM images of heterogeneous samples, computer vision techniques can be applied for feature detection and extraction. Computer vision techniques have been widely used for object identification, medical imaging, satellite image analysis, and numerous other applications. It is a well-established technique applied to labor intensive processes to accelerate identification of objects as well as automate feature extraction. Computer vision assisted techniques have also been utilized in materials science for microstructure characterization and recognition [1]–[3], including powder characterization for additive manufacturing [4]. Computer vision feature detection techniques such as Harris-Laplace [5], Difference of Gaussian [6], Haralick texture features [7], and histogram of oriented gradients [1] have been previously utilized. In particular, the "bag of visual words" image representation employed by Holm et al. [8] to create "fingerprint" microstructures is a good example of using computer vision techniques to extract information from micrograph images. In addition, more recent cognitive neural network based approaches have also been utilized [9] to help identify molecular assemblies on surfaces and microstructures. These previous works and approaches have shown significant values in analyzing visualization data that can significantly increase the throughput of often labor and time-intensive image data analysis.

While prior works have focused on characterization, our work takes computer assisted image processing and analysis a step further, to correlate image features with materials performance. In order to demonstrate this capability, we focus on correlating features from SEM images of organic crystal microstructures (i.e. crystal size, morphology, surface area, etc.) to uniaxially compressed peak stress of consolidated 2,4,6-triamino-1,3,5-trinitrobenzene (TATB) samples, while holding the processing conditions constant. TATB is an insensitive high explosive compound of interest for both Department of Energy and Department of Defense [10]. Although the specific models that are developed in this work only pertain to the use case described herein, the methods and the conclusions that are presented can have a broad implication for materials scientists who can adopt the developed approach for a variety of applications.

## 2. Technical Approach: Computer Vision and Deep Learning

Deep learning (DL) has demonstrated advantages over traditional machine learning (ML) and computer vision (CV) techniques for a variety of applications, most notably: improved predictive performance and automated learning of feature representations with minimal human guidance. However, important limitations remain. In particular, DL typically requires more labeled training examples than traditional ML approaches, and it is often difficult to explain model performance. In order to assess application of computational tools for materials science, we chose to compare two approaches: (1) a traditional ML approach (random forest) using state-of-the-art computer vision features and (2) an end-to-end deep learning approach.

### 2.1 Computer Vision

A wide range of features have been produced by the computer vision and image processing communities that can be used to classify images or perform regression on them. We do not know a priori which of these features will be most useful in performance prediction or physical measurement correlations. Ultimately what is desired is a set of features that are complete (i.e., they capture all materials attributes of interest) and concise (i.e., minimize redundant features). To that end, we chose two complementary state-of-the-art image feature extractors: (1) Bag-of-Visual-Words [11] to capture local shapes and (2) Binarized Statistical Image Features [12] to capture image textures.

A common technique known to perform well for general image classification is the Bag-of-Visual-Words (BoVW) [11] using Scale-Invariant-Feature-Transform (SIFT) vectors [12]. SIFT captures local shapes (i.e., edges, corners, blobs, ridges) and is robust to changes in scale, rotation, illumination, and viewpoint. This technique has been successfully applied to materials science with microstructural image data [1], [2], [8] and we hypothesize that BoVW will capture the relevant microstructure in TATB SEM images as well. The algorithm works by computing SIFT features on all images and then clustering these features using k-means to establish the visual "words." A description vector is then formed for each image by assigning the output SIFT vectors to a cluster and computing the histogram (i.e. how many vectors are in each cluster for a given image).

An extension of BoVW known as Vector-of-Locally-Aggregated-Descriptors (VLAD) [13] utilizes a similar feature encoding pipeline to BoVW but characterizes the distribution within the cluster through the cumulative residuals in each dimension. Expressivity of the feature vector increases as the spatial distribution in each cluster is reflected in comparison to just cluster assignment. This step aims to mitigate the assumption carried over from Bag-of-Words (BoW) that each cluster (i.e., "word") is a single point with zero area. Furthermore, VLAD commonly replaces the k-means clustering algorithm with a Gaussian Mixture Model (GMM) for soft cluster assignment. Soft assignment allows overlap amongst cluster distributions which can be considered during residual calculations for the VLAD encoding. Finally, KAZE [14] is used as a replacement for SIFT due to comparable, if not better, performance in detection and description as well as ease of use in recent versions of OpenCV [15]. The VLAD encoded description vector is created by flattening the set of cluster residuals producing a description vector of length k × d where k is the number of clusters and d is the dimension of the KAZE feature.

With VLAD able to capture local shape information, we turn to image texture features as a way to capture differences in surface appearances (*the suggestion of investigating surface appearances of TATB originated from a subject matter expert in mechanical performances of TATB*). The computer vision literature is full of methods for capturing image texture features [16]. The technique of Binarized Statistical Image Features (BSIF) [17] is a relatively recent and robust algorithm for separating distinct textures. BSIF works by binarizing convolutional responses to pre-learned filters and outputting the responses in a histogram, resulting in a 255-length vector descriptor for each image. The filters are computed by way of independent component analysis on a large set of sample images.

Given VLAD and BSIF features, we use supervised machine learning to train a regression model that can predict materials performance given a corresponding SEM image. Given training SEM images, labeled with known material performance values, the training procedure is: (1) extract VLAD and BSIF features from the image and (2) train a random forest (RF) regressor using these image features. The result of this training procedure is a RF regression model that will output a material performance prediction when given a new SEM image as input. We chose RF as a representative ML algorithm because it requires no meta-parameter tuning and has been shown to perform well for a wide range of ML problems.

### 2.2 Deep Learning

As an alternative to training a traditional ML model (e.g., RF) using a fixed set of extracted CV features, we consider an end-to-end deep learning (DL) solution. Again, we employ a supervised training procedure using labeled examples of SEM images. However, the end-to-end DL solution does not require a separate image feature extraction step. Instead, the DL learning algorithm requires only the raw SEM image pixels (plus the supervised materials performance labels) as inputs. Image features are extracted automatically based on characteristics of the data as part of the DL algorithm. Our DL approach consists of: (1) pretraining a deep convolutional neural network (CNN) on ImageNet [18] data, followed by (2) a supervised training phase using raw SEM images as input.

# 3. Case Study: Predicting Mechanical Performance of Consolidated TATB

## 3.1 Background

Mechanical properties of consolidated molecular solids are important performance criteria for their applications. Molecular solids, such as active pharmaceutical ingredients (API) and high energetic (HE) compounds, are often used in their consolidated forms (i.e. tablets and pressed parts) [19], [20]. Many factors govern the mechanical performance of consolidated parts, including attributes of the starting materials, pressing and processing conditions.  In this work, we focus on the characteristics of the starting crystals (or particles), including crystal size, shape, surface texture (i.e. roughness), porosity and defects, as captured by computer vision and DL features to understand how they are correlated to performance (Figure 1). Aforementioned crystal attributes have been previously shown to impact mechanical performances, therefore conventional measurement techniques such as laser light scattering and Brunauer-Emmett-Teller (BET) measurement are used to capture such information and then used to correlate to performances [21], [22].  However, each of these measurements provide only a fraction of the information about the crystal attributes needed to make strong correlations. We postulate that with recent advancements, both computer vision and DL methods can efficiently capture the crystal attributes mentioned above from SEM images, and have just as good, if not better, correlation to mechanical performances. It is important to note that the sample preparation, pressing and processing conditions are held constant for all samples to tease out only the crystal attributes' contributions to the mechanical performance.

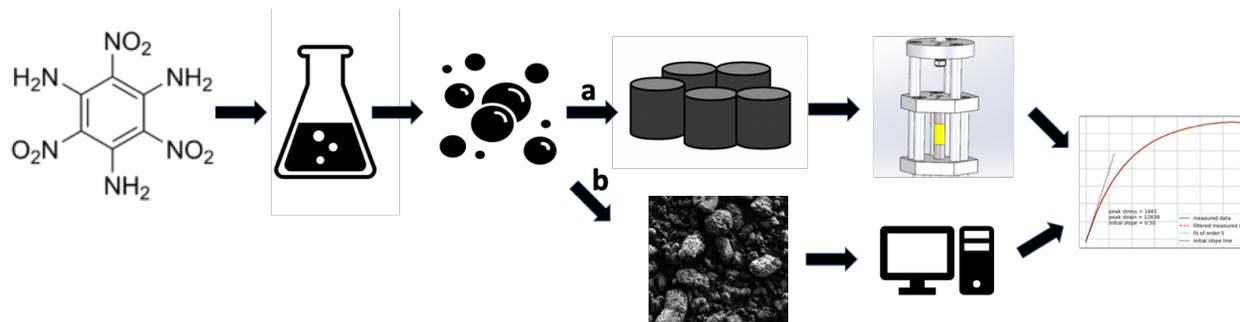

Figure 1. Schematic representation of work flow for a) mechanical property measurement and b) performance prediction using machine learning.

Figure 2 shows typical SEM images of TATB crystals. Depending on the synthesis reaction conditions, different looking TATB crystals can be synthesized. Quantifying TATB crystal characteristics and inferring the significance of these very different looking TATB crystals require significant prior knowledge and experience. To aid in this difficult task, computer vision techniques can be applied to extract image features from SEM images to provide quantifiable TATB crystal features. The extracted features can then be correlated to mechanical performance of consolidated TATB created from different lots of TATB crystals using machine learning (while holding constant the image collection process as well as sample processing and pressing conditions). With a robust regression model correlating TATB features to mechanical performance, one can determine the key TATB features which dominate the mechanical properties of the consolidated TATB parts. This may provide insights to understand the fundamental TATB microstructures that contribute to the mechanical performance of consolidated TATB parts and guide synthesis processes to achieve the desired TATB crystal features.

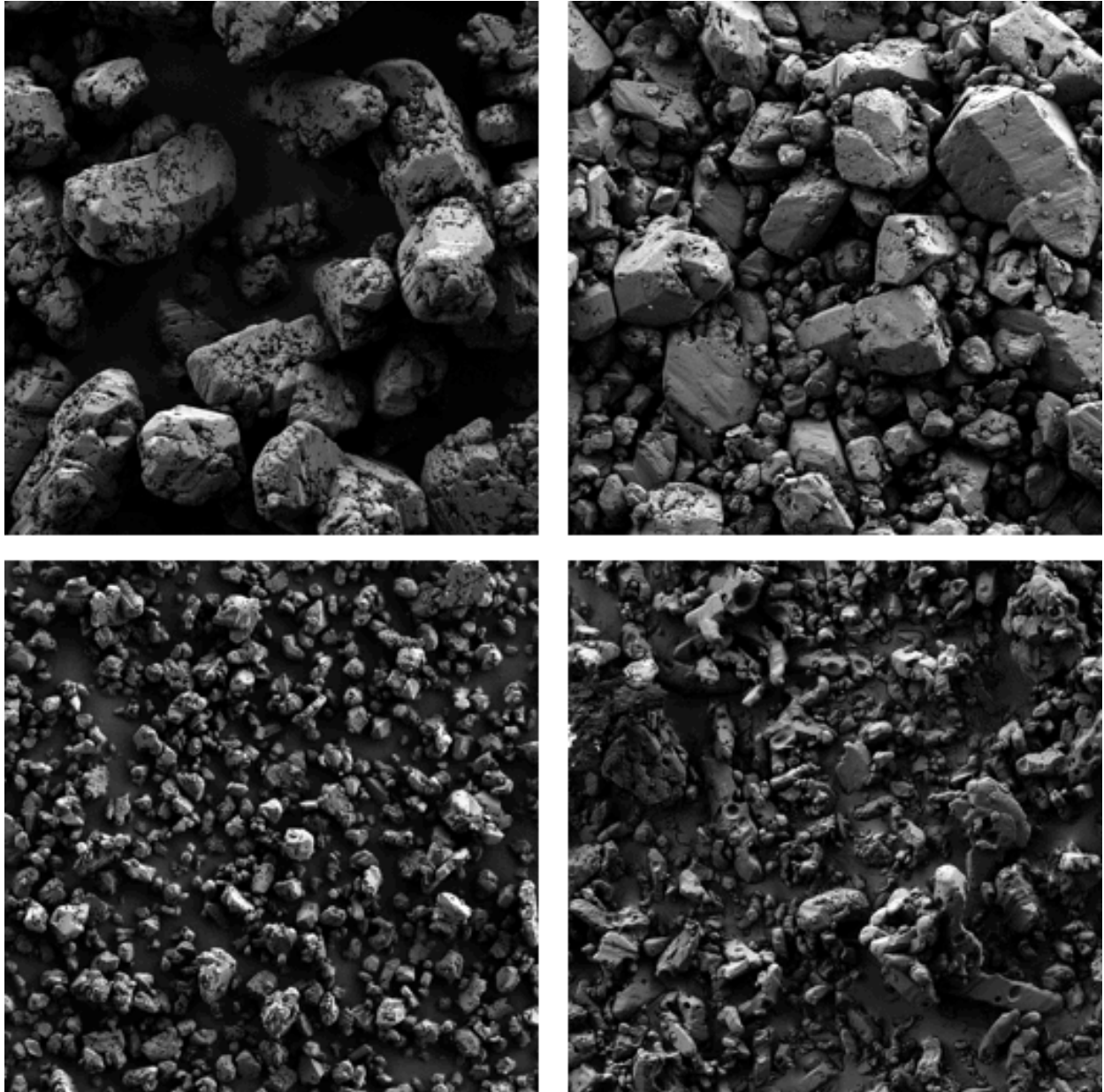

**Figure 2. Examples of different TATB crystal structures with varying synthesis conditions at identical magnifications. The field of view for each image is 256.19 μm x 256.19 μm.**

Herein, we report our efforts to develop a method to predict a figure-of-merit (compressive peak stress) for various lots of TATB, based solely on SEM images, by leveraging computer vision and machine learning techniques.

## 3.2 Methodology

### 3.2.1 Consolidated Material Sample Preparation & Compressive Testing

The TATB lots (a lot refers to TATB crystals produced during 1 synthesis run; each lot is assigned an alphabetical identifier) were selected to span the extremes of the both the compression performance range and the known material feature space (e.g., very small and very large crystallites). Figure 3 shows the "ground truth" compression performance measurements for each lot. These ground truth values are calculated by taking 2 separate measurements on 2 separate samples per lot and averaging the two. Overall, we observe strong agreement between the measurements within each lot (the mean percent difference between the two is 2% and the median is 1%). One notable exception is lot R, for which the 2 measurements vary by 14%. As we observe in Section 3.3, our prediction error on lot R is considerably higher than any of the other lots, perhaps due to this uncertainty in the correct peak stress value. TATB powder from each lot was uniaxially pressed in a cylindrical die at ambient temperature to 0.5 inch diameter by 1 inch height, with a nominal density of 1.800 g cc$^{-1}$. Strain controlled compression tests were run in duplicate at 23°C at a ramp rate of 0.0001 s$^{-1}$ on an MTS Mini-Bionix servohydraulic test system model 858 with a pair of 0.5-inch gauge length extensometers to collect strain data. From the obtained stress-strain curve, only the peak stress values were considered as the outputs of the machine learning models (see Figure 1).

### 3.2.2. Image Data Collection

For capturing microstructure information of TATB using SEM image analysis techniques, it is imperative that sample preparation and collection are consistent. For sample preparation, TATB powder is adhered to SEM stubs using double-stick carbon tape on the stub, that is placed gently into a reservoir of TATB powder. The excess loose TATB powder is gently blown off with compressed air. The samples are coated with nominally 3.3nm of gold prior to imaging.

The SEM images are collected with a Zeiss Sigma HD VP using a 30.00 μm aperture, 2.00 keV beam energy, and *ca.* 5.1 mm working distance. The software *Atlas* is used to automate the image collection. An area of interest is selected by the user (in this case, the entire SEM stub surface) and Atlas subdivides this area and collects zoomed-in images with slight overlap to create a stitched mosaic of the full area.

In this analysis, we are using the individual tiles as our image population. The image tiles are set to have a field of view of 256.19 μm × 256.19 μm with a pixel size of 250.18 nm × 250.18 nm and to autofocus every 20$^{th}$ tile. The images used in this analysis are collected using the SE2 secondary electron detector. The brightness and contrast levels are held constant across all images and samples.

In all, we collected 69,894 sample images from 30 lots of TATB. Each image is associated with the single peak-stress value for the lot, measured mechanically as described in the previous subsection.

### 3.2.3. Machine Learning Implementation

Unless otherwise specified, we use default settings for all software libraries.

We implemented VLAD in Python using OpenCV [15] for the KAZE image descriptor and scikit-learn for GMM clustering. We set k = 20 and d = 64, where k is the number of clusters and d is the dimensionality of the KAZE vector, to keep overall dimensionality low and create a small dictionary, motivated by the homogenous nature of the TATB lots. During the KAZE key point extraction, we consider only the top 128 key points based on the response value, providing 128 KAZE features per image to be fed into the clustering algorithm.

For BSIF, we use the code and pre-computed filters available on the authors' website [23]. We translated the original Matlab code to Python for use within our learning framework.

We train random forest regressors using Python scikit-learn [24] with the following configuration. We use 100 trees, N samples per tree (the original input sample size), but samples are drawn with replacement, a maximum tree depth of 32, a minimum of 1 sample per leaf, and a 1/3 sample of features are considered for each split (standard for random forest regression). Note that the scikit-learn *RandomForestRegressor* default for *max_features* is N, which is actually a degenerate random forest with no feature sampling.

Our DL approach consists of training deep convolutional neural networks (CNNs) using the Python Caffe framework [25]. We started with an ImageNet [18] pretrained DenseNet 121 network [26]. The SEM image is gray-scale, but it is converted to RGB (using the OpenCV cvtColor function) in order to make it compatible with the DenseNet network. The target material performance values are normalized so that they range from approximately -1.0 to 1.0. The network has an input size of 352×352, but the SEM images are scaled to 384×384 using bilinear interpolation (standard OpenCV resizing). This allows us to do random cropping during training. We performed mean subtraction preprocessing using the mean derived from ImageNet, which seems to work better in practice than computing a dataset-specific mean for each new dataset. The initial learning rate is 0.01 using an exponential rate step down. We train for 20,000 iterations with a step size of 200 iterations and a learning rate decay of 0.94 for each step. Mini-batch size is 32. We use standard stochastic gradient descent training with momentum set to 0.9 and weight decay set to 0.0002.

### 3.2.4. Machine Learning Experiments

To evaluate the ability of the proposed machine learning approaches to generalize to previously unseen materials, we employ a leave-one-lot-out cross-validation procedure. For each of the 30 lots *L*: we train a model on all lots other than *L* and then evaluate the trained model on lot *L* only. We use the trained model to predict peak-stress for each image in the evaluation lot and then calculate a single peak stress prediction for the lot as the median prediction over all images in the lot.

## 3.3 Results

### 3.3.1 Peak-stress Prediction: Random Forest vs. Deep Learning

Figures 3 and 4 show respectively the peak stress predictions and mean absolute percentage error (MAPE) for RF and DL on each lot. The error bars in Figure 3 represent standard deviation of predictions across images in a lot. Overall, DL outperforms RF, achieving 206 root mean square error (RMSE)/10% MAPE across all lots vs. 271 RMSE/13% MAPE for RF. However, Figure 4 highlights several exceptions, where RF error is lower than DL. These are lots E, F, AX, AT, V, and AW. We also note the large discrepancy in performance between RF and DL on lot R (as noted in Section 3.2.1, there is uncertainty regarding the correct peak stress for lot R, as the 14% measurement error for lot R is much higher than the 2% average). However, even removing lot R from the evaluation, DL exhibits lower error overall (200 RMSE/9% MAPE vs. 235/11% for RF). Figure 4 also shows the performance of a simple baseline approach that doesn't use the image data at all but makes use of the distribution of peak stress values, in this case by always predicting the mean peak stress value across all lots (1,580 psi). This type of baseline approach is standard practice in the machine learning community and helps differentiate the effects of distributional information available in training labels from the feature information available in training images. This baseline approach achieves 419 RMSE/26% MAPE overall.

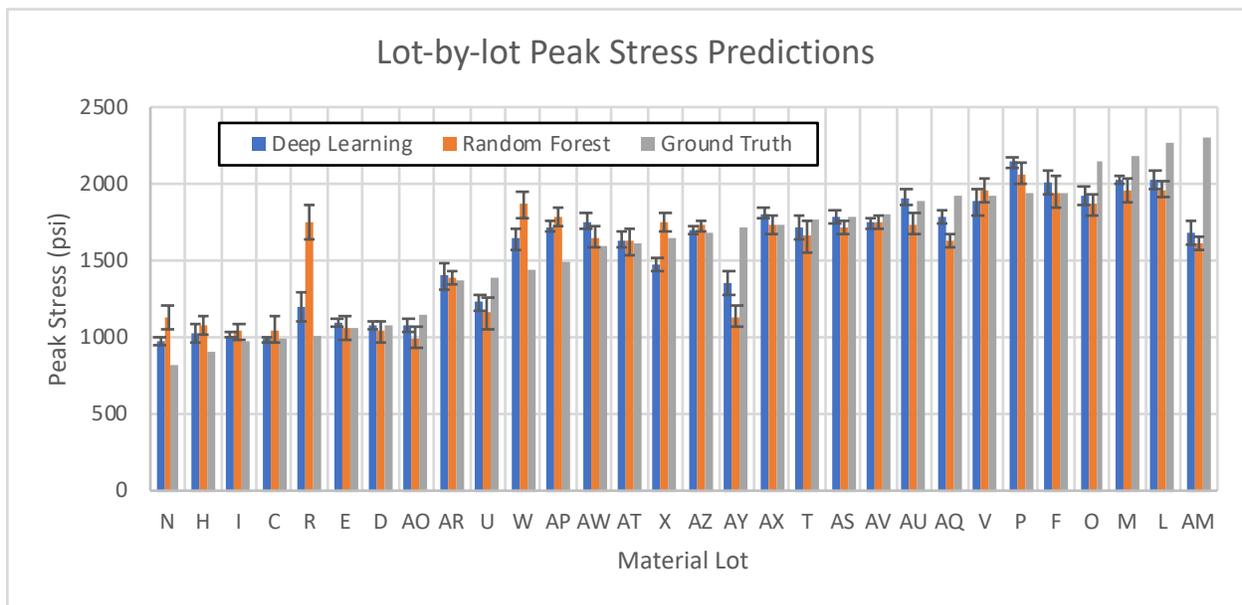

Figure 3: Lot-by-lot predicted peak stress values for both the CV/RF approach and the DL approach, as well as the observed ground truth peak stress values from mechanical testing. Closer to ground truth prediction is better. Lot prediction is the median prediction across all images in the lot. Error bars represent standard deviation of predictions across images in a lot.

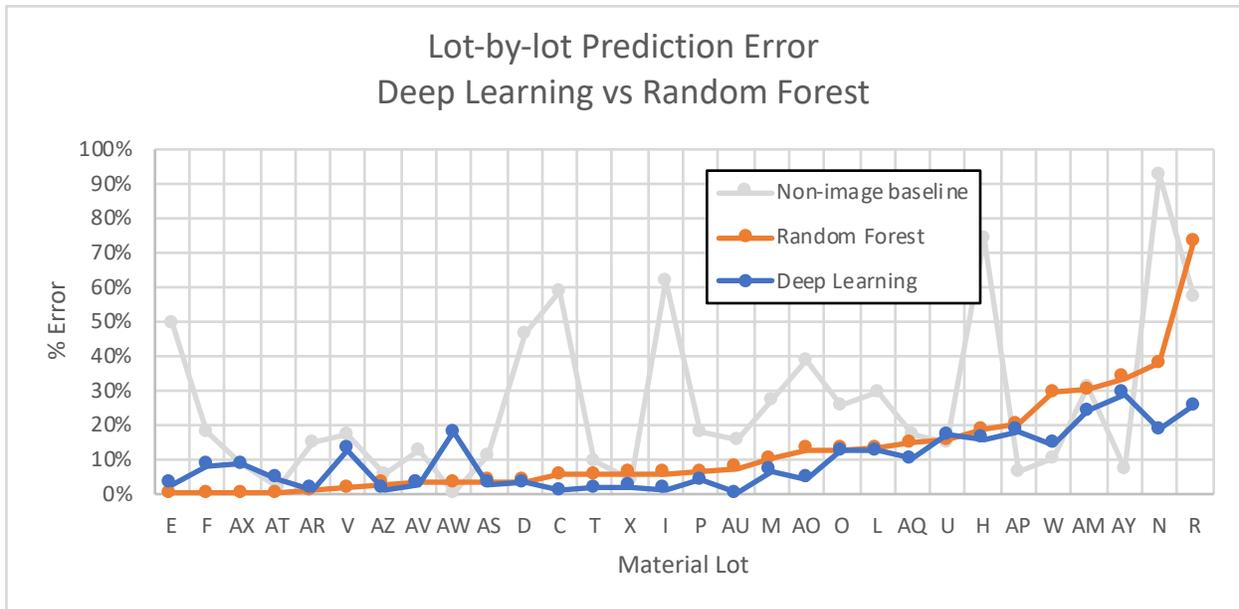

**Figure 4:** Lot-by-lot prediction error for RF, DL, and a non-image baseline (always predict mean peak stress). Error is measured using mean absolute percentage error (MAPE). Lots are ordered by increasing RF error to make differences between RF and DL clearer. Lower error is better.

### 3.3.2 Learning Curves

In order to understand how model performance is affected by the availability of training data, we generated learning curves for RF and DL using the following cross-validation procedure with training set subsampling, which allows us to both: (a) vary the number of training lots available while (b) evaluating on each test lot exactly once:

    For each value of *T* = (5, 7, 9, …, 29):
        For each of 30 lots *L*:
            *L* is the test lot and is excluded from training
            Randomly select additional lots for exclusion until exactly *T* training lots remain
            Train a model M on these *T* training lots
            Test Err = evaluate *M* on held-out test lot *L*
        Report mean Test Error over each lot for *T*

Figure 5 shows two versions of the resulting learning curves: Figure 5a shows the raw data, where each point represents an average over 30 cross-validation folds (as described above). Figure 5b shows a smoothed version of the same plot, transformed by a central moving average filter with a window size of 3. This smoothing is effectively a low-pass filter, which removes high-frequency fluctuations (i.e., trial-to-trial variance) and exposes the underlying trend. For both RF and DL, generalization error drops as we add more training lots. Therefore, it appears that variance (i.e., overfitting) is a significant source error for both models. RF is less subject to overfitting than DL for small training sizes since it is a lower complexity model. However, as more training lots are provided, RF performance plateaus as model capacity is exhausted and RF begins to underfit the data. DL performance continues to improve steadily right up to training on 29 of 30 lots, indicating that even 30 lots is not enough to take advantage of the full DL model capacity. This strongly suggests that the error of the current DL modelling approach will decrease simply by training on more material lots, as they become available.

For this application, the transition point where we have enough data that DL begins to outperform RF occurs at approximately 20 training lots. We note that this transition point is highly data-dependent and does not represent a universal rule. For example, 20 lots of identical material may be no better than 1 lot. Fortunately, in practice, for a given data set, one can determine through cross-validation on the training set whether or not enough data has been collected that DL will outperform RF.

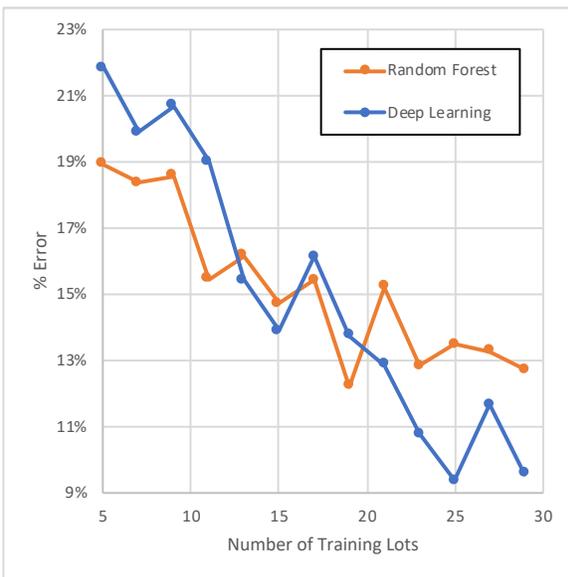
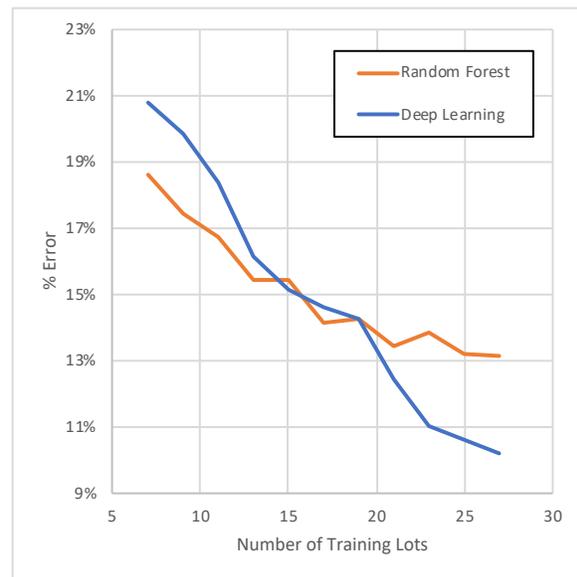

(5a) Raw Curves        (5b) Smoothed Curves

Figure 5: Raw (5a) and smoothed (5b) learning curves for RF and DL. Error is measured using mean absolute percent error (MAPE). DL overfits early on, but the fit gets better and better with more data. RF is less subject to overfitting for small training sizes, but eventually underfits as model capacity is exhausted. Lower error signifies better model performance.

### 3.3.3 Comparison of Computer Vision Feature Sets

Figure 6 shows the percent error (MAPE) for RF models trained using different feature sets on each lot. Overall, BSIF produces a lower-error model (272 RMSE/13% MAPE) than VLAD (329 RMSE/19% MAPE). However, Figure 6 highlights several exceptions, where VLAD error is lower than BSIF. The most notable are lots T, U, AP, W, and AY. Combining BSIF with VLAD does not significantly improve performance beyond BSIF on its own. This suggest that particles' texture (captured in images by BSIF) is a stronger indicator of the compressed materials' bulk compressive strength than particles' local shape (captured by VLAD).

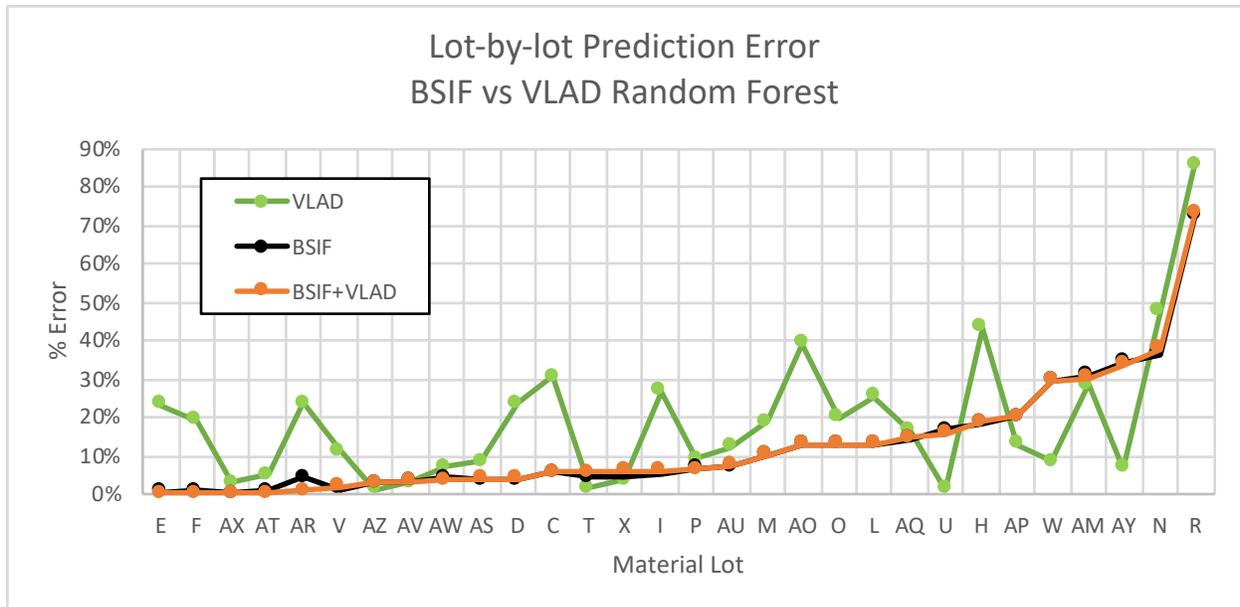

**Figure 6: Lot-by-lot prediction error for RF trained on different feature sets: VLAD only, BSIF only, and a combined model using both BSIF and VLAD. Error is measured using mean absolute percentage error (MAPE). Lots are ordered by increasing BSIF error to make differences between feature sets clearer. Lower error is better.**

### 3.3.4 Comparison with State-of-the-Practice Baselines

In addition to the simple statistical baseline described in Section 3.3.1 (Figure 3), we consider two domain-specific baselines that are more representative of the current state of the practice: (1) a human assessment of material microstructure, based on SEM images and (2) particle size distribution (PSD) and surface area measurements from laser scattering and BET instruments.

Human assessment features (HAF) were generated by two human domain experts, based on a visual inspection of approximately 10 SEM images from each lot. The features considered were: particle size, porosity, polydispersity, facetness, and surface area. Each feature was assigned a relative value on a scale from 0 to 1, based on the subjective assessment of each human expert. For each feature and lot, the two human assessed values were averaged to obtain a final feature value. We also calculated the standard deviation of the two assessed values as a measure of agreement between the human experts. Across all features and lots, the average standard deviation was 0.11, indicating a reasonable consensus among experts.

Some attributes of TATB particles can be measured directly by instrumentation. We can evaluate these attributes as predictors of the consolidated materials' compressive strength. These instrument-measured features provide another point of comparison for the predictive performance of our CV-derived features. Laser scattering based particle size distribution measurements were obtained with Malvern Mastersizer 3000 and BET surface area measurements were made with Micromeritic Tristar II. The BET analysis resulted in 1 scalar value feature for each lot signifying surface area of the materials in g/cc and PSD analysis resulted in 6 scalar values as features which describes crystal size distribution in volume percent in 6 different size ranges (<15 μm, 15-35 μm, 35-60 μm, 60-85 μm, 85-105 μm and >

105 μm). Due to sample quantity availability, we obtained PSD/BET values for 17 samples of the 30 total lots (see Table 1). Therefore, the evaluation described here is a leave-one-lot-out cross validation over these 17 lots only. We report average MAPE across lots.

Table 1. BET and PSD measurements of 17 lots of TATB. PSD numbers are reported in volume %.

| Lot | BET (g/cc) | <15 micron | 15-35 micron | 35-60 micron | 60-85 micron | 85-105 micron | >105 micron |
|---|---|---|---|---|---|---|---|
| AM | 0.880 | 0.000 | 9.920 | 32.490 | 35.580 | 16.540 | 5.470 |
| AO | 0.344 | 25.770 | 40.720 | 16.040 | 8.230 | 4.070 | 5.170 |
| AP | 0.421 | 11.286 | 9.451 | 16.397 | 31.878 | 21.374 | 9.614 |
| AR | 0.511 | 64.465 | 25.385 | 3.740 | 4.710 | 1.165 | 0.535 |
| AS | 0.528 | 8.280 | 23.035 | 34.565 | 22.150 | 7.820 | 4.150 |
| AV | 0.498 | 25.365 | 55.660 | 16.760 | 2.185 | 0.030 | 0.000 |
| AW | 0.495 | 39.640 | 37.320 | 16.340 | 3.790 | 1.900 | 1.010 |
| AZ | 0.417 | 8.695 | 28.055 | 33.415 | 19.385 | 6.545 | 3.905 |
| E | 0.277 | 8.836 | 16.111 | 21.937 | 22.018 | 15.528 | 15.570 |
| H | 0.337 | 9.450 | 25.440 | 33.820 | 20.050 | 7.560 | 3.680 |
| M | 0.764 | 44.600 | 44.580 | 9.710 | 0.600 | 0.060 | 0.450 |
| N | 0.220 | 1.730 | 2.070 | 28.170 | 34.340 | 17.620 | 16.070 |
| P | 0.493 | 20.100 | 42.510 | 26.150 | 8.890 | 2.170 | 0.180 |
| R | 0.380 | 51.380 | 38.580 | 7.120 | 1.660 | 0.880 | 0.380 |
| V | 0.683 | 68.210 | 24.610 | 4.320 | 1.720 | 0.680 | 0.460 |
| W | 0.404 | 11.710 | 16.150 | 23.590 | 19.760 | 11.060 | 17.730 |
| X | 0.501 | 15.980 | 16.710 | 25.660 | 20.010 | 10.090 | 11.550 |

Figure 7 presents a comparison between our computer vision approach (a RF model trained on BSIF features) and three baseline RF models trained on various combinations of the human assessment and PSD/BET features. The figure shows that the computer vision approach outperforms the baseline approaches by a significant margin, indicating that BSIF captures some additional information that is contained in neither the human assessment nor PSD/BET measurements. Interestingly, training a RF model on both BSIF and baseline features together performs no better than BSIF on its own, indicating that BSIF is effectively capturing all of the information provided by the human assessment and the PSD/BET instruments, as well as some additional information.

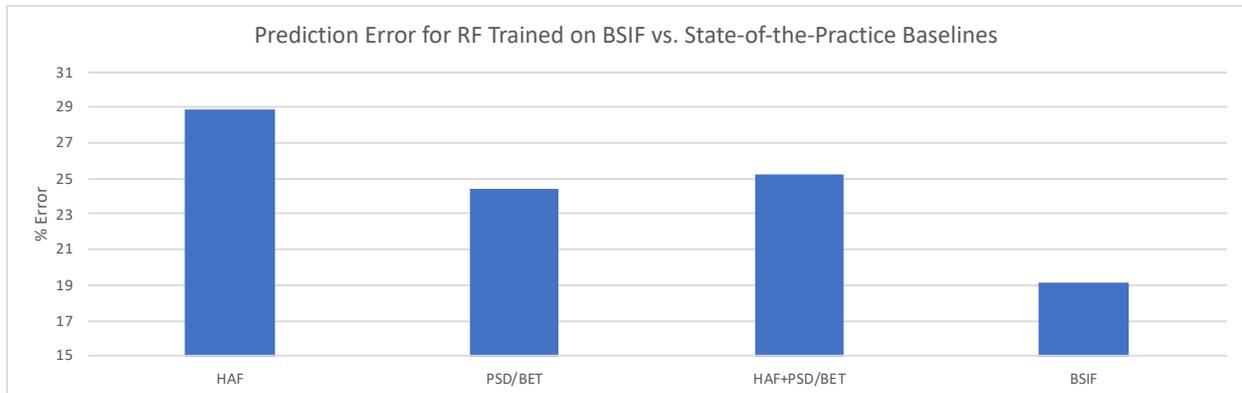

Figure 7: Comparison of RF models trained on computer vision (BSIF) features vs. baseline features representative of the current state-of-the-practice (human assessment features (HAF) and PSD/BET). The results indicate that the computer vision approach is able to capture the information captured by both domain-expert assessments and the PSD/BET instruments, plus some additional information.

### 3.3.5 Explaining Model Performance

As discussed in the previous section, the proposed computer vision models can capture important material attributes from SEM images that are not adequately captured by subjective domain expert assessment. What the above results do not tell us is: what specifically are these new features? One general approach for determining what image features correlate with material performance is to remove specific features from the images and measure the effects of this removal on model performance. If we remove a feature *F* and the model performance predictions change significantly, then clearly *F* is an important indicator of material performance. Here, we apply this general technique to determine whether the model is keying in on finer or coarser image features. Specifically, we wish to evaluate two hypotheses:
  A. The presence of fine crystal attributes (such as pores and defects) in TATB crystals indicates a higher strength material (i.e., higher compressive peak stress).
  B. Our predictive models rely on these fine crystal attributes (as captured by BSIF texture features) for accurate peak stress prediction.

To evaluate hypothesis A, we demonstrate that material lots containing fewer fine attributes underperform compared to material lots that contain more fine attributes. To evaluate hypothesis B, we artificially remove fine features from SEM images and demonstrate that this results in lower predicted performance from our models.

First, we observe that changes to the synthesis protocol produce variation in the presence of fine crystal attributes. Figure 8 provides a qualitative comparison of a lot with abundant fine attributes (Lot AM) to a lot with fewer fine attributes (Lot N). Next we describe a procedure for removing fine features from existing SEM images as well as a measure for quantifying the degree to which a material lot contains these fine attributes to begin with.

To remove fine features from SEM images, we apply a bilateral "blurring" filter, which takes a gaussian-weighted average of neighboring pixels while also considering intensity differentials. The overall effect is that fine features are removed while preserving stronger edges, such as crystal boundaries. To implement this filtering we use the *bilateralFilter* function from OpenCV [15] with *sigmaColor* = 75 and *sigmaSpace* = 75. We control the amount of blurring by varying the filter size from 6x6 pixels up to 30x30 pixels. A small filter removes only the finest features, whereas larger filters remove increasingly coarser features. Figure 9 demonstrates the effects of this blurring filter on material Lot C.

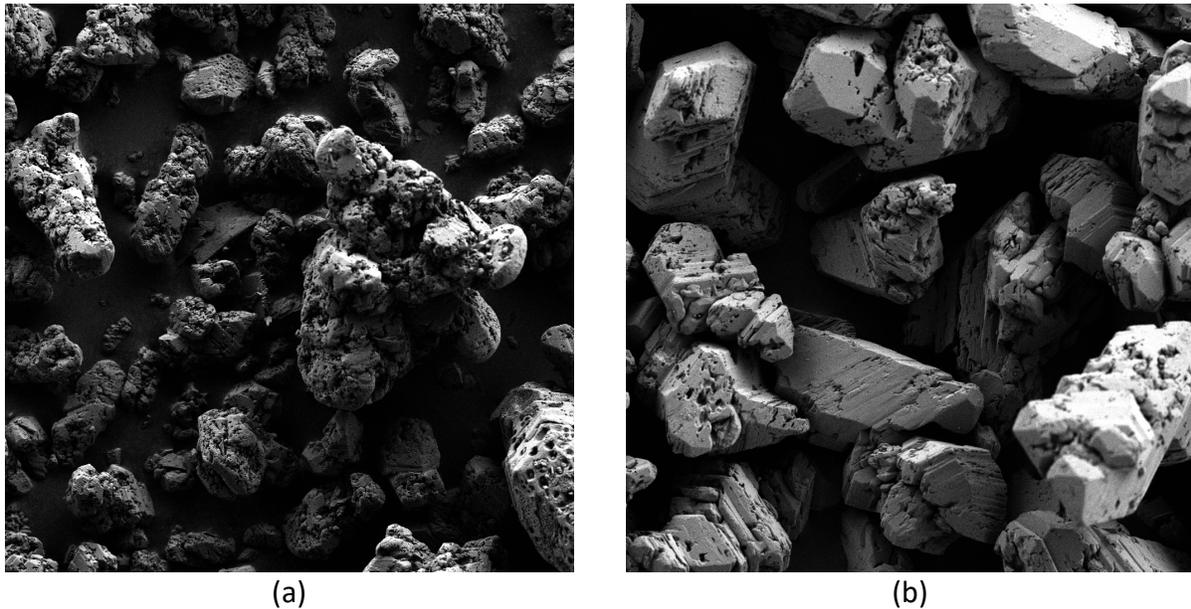

(a) (b)

Figure 8: Material lots differ in terms of the presence of fine crystal attributes: (a) Lot AM crystals contain abundant fine attributes, (b) Lot N crystals contain fewer fine attributes.

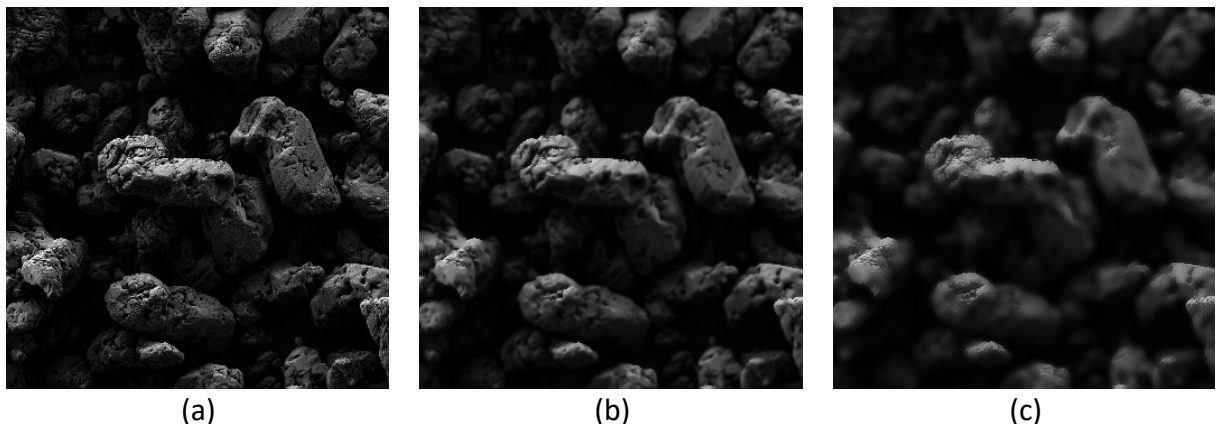

(a) (b) (c)

Figure 9: The bilateral blurring filter removes fine features from images: (a) unfiltered image from Lot C, (b) image blurred with 12x12 pixel filter, (c) image blurred with 30x30 pixel filter.

Utilizing this blurring filter, we conduct an experiment to evaluate hypothesis B: that our predictive models rely on fine crystal attributes (as captured by BSIF texture features) for accurate peak stress prediction. To do this, we repeat our leave-on-lot-out evaluation protocol for Random Forest, with one modification: we apply blurring to the images in the evaluation lot to remove the fine features, varying the size of the blurring filters for each experiment. Figure

10a shows the results of these experiments. When fine features (i.e., 12x12 pixels or smaller) are removed from high performance lots P, M, and AP, our models predict lower performance on these lots than when fine features are present. However, low performance lots E and C are largely unaffected by removal of fine features, since they contain few fine features to begin with.

Next we evaluate hypothesis A: that the presence of fine crystal attributes in TATB powder indicates a higher strength material (i.e., higher compressive peak stress). To do this, we define a measure to capture the presence of fine image features based on the blurring procedure described above:

$$fineness(S) = \frac{\sum_i |S_i - B_i|}{\sum_i S_i}$$

where $S_i$ is the $i^{th}$ pixel of the original SEM image S and $B_i$ is the $i^{th}$ pixel of the blurred image.

The intuition behind this measure is that the more blurring changes an image, the more fine features must have been naturally present in that image (higher *fineness*). If blurring has little effect on an image, then the image contained few fine features to begin with (lower *fineness*).

Using this *fineness* measure (with a 30x30 pixel blurring filter), we assess the degree to which each material lot contains fine features and correlate this with the performance (compressive peak stress) of each lot. Figure 10b shows a clear correlation between the presence of fine features and material performance. Note that this experiment does not depend on our predictive model in any way. We are simply showing that material lots that naturally contain fewer fine features underperform material lots that contain more such fine features.

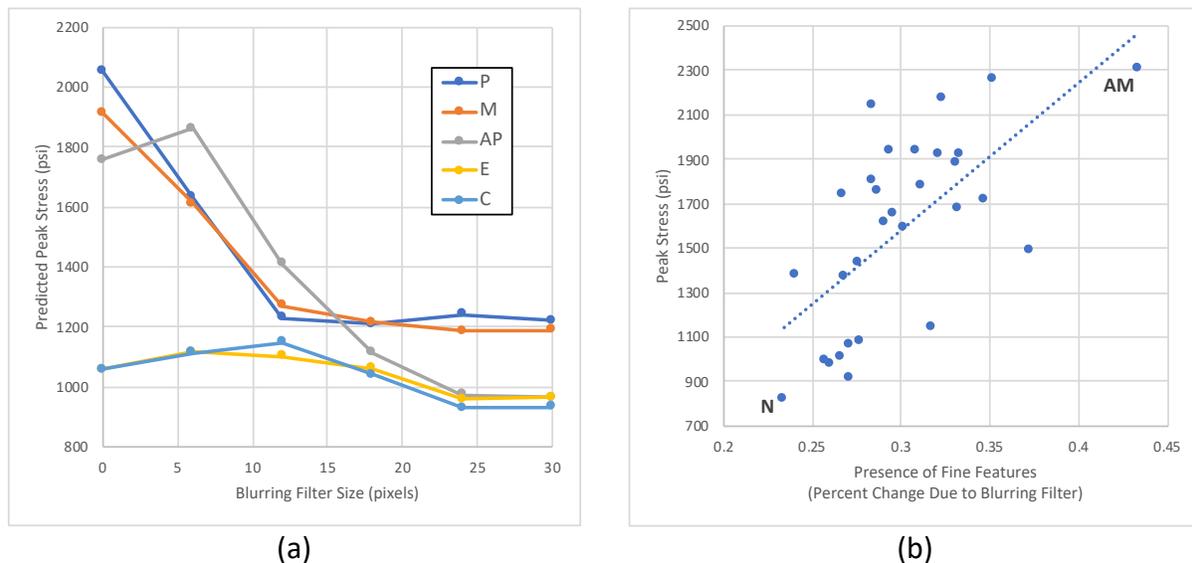

(a)      (b)

Figure 10: (a) When fine features are removed from high performance lots P, M, and AP, our models predict lower performance on these lots. Low performance lots E and C are largely unaffected by removal of fine features, since they contain few fine features to begin with. (b) Material lots that naturally contain fewer fine features underperform material lots that contain more such fine features. Here, we calculate fineness using a 30x30 pixel blurring filter.

### 3.4 Discussion

**Computer vision is an effective approach for correlating material microstructure with performance (provided that the processing condition remains constant)**. Figures 3 and 4 show that: (a) the SEM images do in fact contain information that is correlated with material performance and (b) both engineered computer vision features and automatically learned DL features are effective in extracting this information from the images. The baseline error for this task is 419 RMSE/26% MAPE, achieved by a simple approach which ignores image data completely and makes use of only the distribution of peak stress values. Compared to this baseline, RF reduces RMSE by 35% and MAPE by 50% on average, and DL reduces RMSE by 51% and MAPE by 62% on average. Therefore, there is clearly signals in the image data that both RF and DL are able to effectively correlate with peak stress. Furthermore, Figure 6 demonstrates domain expert guided feature selection (focusing on surface appearance using BSIF) provides better model performance compared to all-inclusive CV features (VLAD). This result highlights the importance of incorporating domain knowledge to model when available. Figure 7 demonstrates that the computer vision approach outperforms current state-of-the-practice (i.e. domain expert led correlation), reducing MAPE by 24% on average. These results indicate that the computer vision approach is able to capture the information captured by both domain-expert assessments and a PSD/BET instrument, plus some additional information. It's important to note that PSD and BET measurements were specifically chosen and used in this experiment because PSD and BET provides similar information CV features are hypothesized to capture. Other materials characterization processes such as computed tomography (CT) will provide additional information computer vision analysis of SEM images would not be able to capture, which may be more relevant. We are currently collecting CT images of consolidated samples of TATB for further analysis.

**Synthesizing more material lots improves performance.** Figure 5 shows that both RF and DL achieve lower generalization error by training over a diverse set of lots. The performance of DL, in particular, improves sharply as more training lots are added. The results indicate that this trend will continue as more lots are collected. Therefore, we expect the error of the current DL modelling approach to decrease simply by training on more material lots, as they become available.

**Deep Learning is the more powerful method.** In addition to the convenience and robustness of DL's automated image feature extraction, we have demonstrated that DL is the best performer overall. Specifically, Figures 4 and 5 show that: (a) DL outperforms RF given sufficient training data (≈20 material lots or more) and (b) the performance gap between RF and DL increases with the number of training lots available. Since DL is a higher-complexity model, it is able to fit whatever data is provided (at least up to 30 lots), whereas RF performance begins to plateau around 15 training lots due to underfitting. However, DL models currently operate as a "black box" approach, where the inner working of the model is difficult to explain. On the other hand, RF model trained on domain expert guided feature selection (BSIF) provides a level of explainability (i.e. surface appearance) which is often a desired outcome of such analysis.

**The more powerful method is not always the best.** The one area where RF consistently outperforms DL is in the "small data" regime, where training lots are scarce (see Figure 5). This is a crucial caveat to the dominance of DL because scientific applications often fall within this small data regime due to the time, effort, and expense required to conduct experiments. It's important note that these two approaches are complementary in their strengths. Our goal is not to declare a winner in a battle of ML algorithms. Rather, our goal is: (1) to illustrate that our overall approach is compatible with any number of ML algorithms and (2) to explore the trade-offs between ML algorithms used for our application. Specifically, there are trade-offs between a high complexity, data hungry, black box model (DL) and a lower complexity, data efficient, more transparent approach (RF). Our expectation is that the DL approach will outperform RF given enough data. However, since material scientists often work with limited data, an empirical study is required to assess whether the available data is sufficient for DL, or instead suggests choosing a simpler model in RF, which may have a limited performance ceiling. Our results provide an important reminder to always compare to simpler methods as baselines, especially when data is scarce. The more powerful method is not always the best.

**Fine crystal attributes are strong indicators of material strength.** Our results show that changes to the synthesis protocol produce differences in the presence of fine crystal attributes, such as pores and defects (Figure 8). These differences correlate with material performance in terms of compressive peak stress (Figure 10b). Finally, our trained machine learning models rely on corresponding fine SEM image features as indicators of material performance (Figure 10a). Note that it is the finest features here which provide strong indicators of material performance (the largest performance drop comes when we move from a 6x6 to 12x12 pixel filter). This may explain why the human assessed features such as porosity and facetness underperform the BSIF features, since the human assessments are more global and cannot adequately capture differences in the finest crystal attributes.

**Future work.** The above discussion suggests a number of avenues for further improving model performance for this task: (1) We expect synthesis of more material lots to automatically lead to gains in DL performance without any changes to the modelling approach. At what point DL performance eventually plateaus is a question for further investigation. (2) In addition to performance prediction, an important goal for experimentalists is extracting insights from machine learned models. We want to understand what specific characteristics of the material microstructure are contributing to performance and, ultimately, how these characteristics are influenced by synthesis parameters. We learned through using VLAD and BSIF features that crystal morphology, porosity and roughness which are readily captured by BSIF features are more important than crystal sizes, which are captured by VLAD features in predicting peak stress of consolidated TATB crystals. Understanding the features extracted by our DL model is an active area of research that is on-going. We plan to investigate materials-specific approaches to explainability. (3) Finally, since our results indicate that DL performance will continue to improve with more data, but more data is sometimes impossible or impractical to obtain, we will investigate methods for augmenting SEM images with other data sources (i.e., particle size analysis, surface area measurements, etc.) as well as artificially generated SEM images.

# 4. Conclusion

Rapid advancements in computer science tools are changing the landscape of data science. Application of computer vision, machine learning, and deep learning in materials science can provide powerful tools to analyze, query and automate scientific data analysis.  As scientific capabilities progress and generate large amounts of data, advanced data analytics tools must be implemented. To that end, we explored the application of computer vision and machine learning to quantify materials properties (i.e., compressive strength) based on SEM images of materials microstructure. We showed that it is possible to train machine learning models to predict materials performance based on SEM images alone, demonstrating this capability on the real-world problem of predicting uniaxially compressed peak stress of consolidated TATB samples. Our image-based ML approach reduces mean absolute percentage error (MAPE) by an average of 24% over baselines representative of the current state-of-the-practice (i.e., domain-expert SEM assessment and surface area and particle size measurements).

We explored two complementary approaches to this problem: (1) a traditional machine learning approach (random forest) using state-of-the-art computer vision features and (2) an end-to-end deep learning approach, where features are learned automatically from raw images. We demonstrated the complementarity of these approaches, showing that random forest performs best in the "small data" regime in which many real-world scientific applications reside, whereas deep learning outpaces random forest in the "big data" regime, where abundant training samples are available.

Finally, we demonstrate that models trained using machine learning techniques are capable of discovering and utilizing informative crystal attributes previously underutilized by domain experts. In particular, we show that fine SEM image features (e.g., smaller than 12x12 pixels), corresponding to fine crystal attributes, provide strong indicators of material performance, which are not adequately captured by subjective domain expert assessment of SEM images.

Based on our findings, we outlined several future research directions in order to further improve the utility of the approach. These include: (1) synthesizing new material lots to better understand the performance ceiling of the deep learning approach, (2) exploring questions of model explainability to extract experimental insights from trained models, and (3) data augmentation approaches to overcome the limited availability of materials samples.

# 5. Acknowledgements
This work was performed under the auspices of the U.S. Department of Energy by Lawrence Livermore National Laboratory under Contract DE-AC52-07NA27344 and was supported by the LLNL-LDRD Program under Project No. 19-SI-001. Authors would like to thank Alan DeHope and Laura Kegelmeyer for providing valuable discussions and information.

## 6. Data Availability

The raw/processed data required to reproduce these findings cannot be shared at this time as the data also forms part of an ongoing study.